\documentclass[10pt,pre,showpacs,preprintnumbers,amsmath,amssymb]{revtex4}
\usepackage{amssymb}
\usepackage{latexsym}
\usepackage{epsfig}
\begin{document}
\title{Thermodynamic descriptions of Polytropic gas and its viscous type as the dark energy candidates}
\author{H. Moradpour$^1$\footnote{h.moradpour@riaam.ac.ir}, M. T. Mohammadi Sabet$^2$\footnote{mohamadisabet@shirazu.ac.ir}}
\address{$^1$ Research Institute for Astronomy and Astrophysics of Maragha
(RIAAM), P.O. Box 55134-441, Maragha, Iran\\
$^2$ Basic Science Faculty, Physics Department, Ilam University,
P. O. Box, 69315-516, Ilam, Iran}

\begin{abstract}
In this paper, at first, we focus on a FRW universe in which the
dark energy candidate satisfies the Polytropic equation of state and
study thermodynamics of dark energy. Bearing the thermal fluctuation
theorem in mind, we establish a relation between the thermal
fluctuation of system and mutual interaction between the dark energy
and dark matter. Generalization to a viscous Polytropic gas is also
investigated. We point to a condition for decaying dark energy
candidate into the dark matter needed for alleviating coincidence
problem. The effects of dark energy candidates and their
interactions with other parts of cosmos on the horizon entropy as
well as the second law of thermodynamics are also addressed. Our
study signals us to a correction term besides the Bekenstein entropy
which carries the information of the dark energy candidate, its
interaction with other parts of cosmos and its viscosity.
\end{abstract}
 \maketitle

\section{Introduction}
The dark sectors of cosmos, at least including dark energy and
dark matter, are mysterious puzzles for the current theoretical
physics \cite{roos}. Among these sectors, dark energy is assumed
as the generator of the current accelerating phase
\cite{Rie,Rie1,Rie2,Rie3}, and in fact, there are various models
introduced for getting a description for the nature of dark energy
\cite{Rev3,Rev1,Rev2}. It seems that observations permit an
interaction between dark sectors of cosmos
\cite{z1,z2,z3,pavonz,z4,z5,z6,z7,z8,wangpavon,z9,co1,z10,co2,co3}
which increases the hopes to solve the coincidence problem if it
leads to decay dark energy into the dark matter
\cite{z1,z2,z3,pavonz,co1,z10,co2,co3}.

The mutual interaction between the dark sectors of cosmos may lead
to thermal fluctuations into the thermodynamic properties of dark
energy candidate, and therefore, one may get expressions for the
mutual interaction between the dark sectors of cosmos in various
models by investigating the thermal history of universe
\cite{wangpavon,fl1,fl2,fl3,fl4,fl5}. Based on this approach, the
entropy of interacting dark energy can be expressed as a function of
entropy of non-interacting dark energy and its derivatives. The root
of this approach comes back to the thermal fluctuation theorem
\cite{landau}, and indeed, thermal fluctuations theorem is valid in
all area of physics such as gravitational systems \cite{landau,das}.
It is also shown that this view leads to find thermal fluctuations
of a Chaplygin gas model of dark energy, by knowing its mutual
interaction form with dark matter, up to the desired order of
thermal fluctuations \cite{emint}. Finally, we should note here
that, in this approach, it is assumed that dark energy candidate
satisfies the first law of thermodynamics and moreover, since the
major part of cosmos is dark, non-dark sectors of cosmos are
neglected from the Friedmann equations
\cite{wangpavon,fl1,fl2,fl3,fl4,fl5,emint}. In addition, the
validity of the generalized second law of thermodynamics in a
universe in which dark energy candidate interacts with radiation and
dark matter is investigated in Ref.~\cite{sara}.

In another approach, by applying the unified first law of
thermodynamics on the trapping or apparent horizon of FRW
universe, some authors show that ghost dark energy model of dark
energy and its generalization may modify the Bekenstein entropy in
the Einstein general relativity framework \cite{cana,cana1}. More
attempts, in which the effects of various models of dark energy
and their interactions with other parts of cosmos on the horizon
entropy are investigated, can be found in Refs.~\cite{em,mmg,md}.
Such modifications also affect the quality of validity of the
second law of thermodynamics and its generalized form
\cite{em,mmg,md}. The unknown origin of dark energy candidate is
the backbone of this approach permitting authors to consider the
possibility of affecting the horizon entropy by the dark energy
candidate. Since all sources which constitute the baryon content
of the cosmos have contributions to the horizon entropy, unlike
the previous approach, one should consider all possible baryonic
sources which fill cosmos \cite{em,cana1}.

Nowadays, the possibility of modelling the universe expansion
history by introducing a candidate for dominated fluid, which
satisfies
\begin{equation}\label{EOS}
p=K\rho^{1+\frac{1}{n}},
\end{equation}
where $p$ and $\rho$ are the corresponding pressure and energy
density, respectively, is investigated \cite{ms}. In this equation
$n$ and $K$ denote the Polytropic index and Polytropic constant,
respectively. It is also shown that, even in the absence of a
cosmic background, such fluid may showcase a transition from a
pressureless state to a state with $\omega=\frac{p}{\rho}=-1$,
where $\omega$ is called the state parameter, and vice versa
\cite{mae}. It seems that its modification and viscous forms may
also use to describe the universe expansion
\cite{phc1,phc2,phc3,hsa}. Indeed, this equation of state, known
as the Polytropic equation of state, is too familiar in
thermodynamics and also for astrophysicists \cite{CALEN,cris}. In
addition, some authors make use of the Polytropic equation of
state to describe dark energy and thus the current phase of
universe expansion as well as to study the thermodynamics of
universe in various theories of gravity
\cite{in,nK,gref1,ref56,karami1,karami2,ref58,ref57,salti}. More
studies, including the numerous properties of a universe in which
the Polytropic gas and its various possible modified forms (as the
candidates of dark energy) interact with dark matter can be found
in
\cite{ahep,setare,ref59,karami4,ansari,karami3,taji,sarkar,ahep1,karami2010}.

Based on the above discussions, it is worthy to study the
thermodynamics of mutual interaction between the Polytropic gas
and its viscous form, as the dark energy candidates, and dark
matter in cosmos in order to establish a relation between the
interaction term and the thermal history of universe by
considering the thermal fluctuation theorem. Moreover, it is also
valuable to study the effects of the dark energy candidate and its
interaction with other parts of cosmos on the horizon entropy and
availability of the second law of thermodynamics.

The paper is organized as follows. In the next section, we
consider a universe filled by dark matter and a dark energy which
satisfies the Polytropic equation of state and find a relation for
the entropy of dark energy, whenever the dark sectors do not
interact with each other. In continue, we generalize our study to
a case in which dark sectors interacts mutually, and by taking
into account the thermal fluctuation theorem, we establish a
relation between the mutual interaction and thermal fluctuation of
system. Generalization to a universe in which a viscous Polytropic
gas plays the role of dark energy candidate is presented in
section ($\textmd{III}$). Finally, we derive a condition for
mutual interaction between dark sectors which is independent of
the equation of state of the dark energy candidate, and should be
obeyed by coupling constants of interaction in order to decay dark
energy to dark matter. Section ($\textmd{IV}$), includes the
effects of a Polytropic dark energy model and its interaction with
other parts of cosmos on the horizon entropy which leads to a
correction term for the horizon entropy. In addition, the quality
of validity of the second law of thermodynamics is also
investigated. Moreover, we also study the effects of a viscous
Polytropic dark energy model and its mutual interaction with other
parts of cosmos on the horizon entropy in the forth chapter. Our
study points two new terms besides the Bekenstein entropy due to
the Polytropic dark energy model and its viscosity. The validity
of the second law of thermodynamics is also investigated for this
model. Section ($\textmd{V}$) is devoted to a summary and
concluding remarks. Throughout this paper we set $c=G=\hbar=1$ for
sake of simplicity.

\section{Thermodynamic description of mutual interaction between non-viscous Polytropic gas and dark matter}
Consider a FRW universe with metric
\begin{eqnarray}\label{frw}
ds^{2}=dt^{2}-a^{2}\left( t\right) \left[ \frac{dr^{2}}{1-kr^{2}}%
+r^{2}d\Omega ^{2}\right],
\end{eqnarray}
where $a(t)$ and $k$ denote the scale factor and the curvature
parameter, respectively. In this metric, $k=-1,0,1$ denotes the
open, flat and closed universes, respectively \cite{roos}. The
apparent horizon of the FRW universe, as the marginally trapped
surface which plays the role of the causal boundary, is evaluated by
\begin{eqnarray}\label{ah2}
\partial_{\alpha}\zeta\partial^{\alpha}\zeta=0\rightarrow r_A,
\end{eqnarray}
where $\zeta=a(t)r$ \cite{Hay2,Hay22,Bak}. By simple calculations,
One gets
\begin{eqnarray}\label{ah}
\tilde{r}_A=a(t)r_A=\frac{1}{\sqrt{H^2+\frac{k}{a(t)^2}}},
\end{eqnarray}
for the apparent horizon radii of the FRW universe
\cite{Hay2,Hay22,Bak,sheyw1,sheyw2}.

Since the major part of cosmos is dark, including dark energy and
dark matter \cite{roos}, we may write the Friedman equation as
\begin{equation}\label{friedman_eq}
H^2=\frac{8\pi}{3}(\rho_m+\rho_D),
\end{equation}
where $\rho_m$ and $\rho_D$ denote the density profile of a
pressureless dark matter and a Polytropic gas as the dark energy
candidate, respectively. We also consider a flat FRW universe in
which $k=0$ \cite{roos}. In a universe, in which the dark sectors
do not interact with each other, the energy-momentum conservation
law implies
\begin{equation}\label{EM1}
\dot{\rho}_m+3H\rho_m=0
\end{equation}
and
\begin{equation}\label{EM2}
\dot{\rho}_D+3H(\rho_D+p_D)=0,
\end{equation}
where $p_D$ is the dark energy pressure, and follows
Eq.~(\ref{EOS}). Defining $\Omega_i=\frac{\rho_i}{\rho_c}$ (where
$\rho_c=3\frac{H^2}{8\pi}$ and $\rho_i$ refers to $\rho_m$ and
$\rho_D$) as a dimensionless parameter, we have the following
equation as the Friedman equation
\begin{equation}\label{friedman_eq2}
\Omega_D+\Omega_m=1.
\end{equation}
For future calculation we also define $u=\frac{\rho_m}{\rho_D}$
leading to
\begin{equation}\label{eq_u}
u=\frac{1-\Omega_D}{\Omega_D},
\end{equation}
where we have used Eq.~(\ref{friedman_eq2}) to obtain this
equation.
\subsection{Thermodynamic description of non-interacting Polytropic gas}
Here, we are going to find a relation for the entropy changes of a
dark energy candidate, obeying the Polytropic equation of
state~(\ref{EOS}). In order to achieve this goal, taking into
account the view in which dark energy candidate satisfies the
first law of thermodynamics
\begin{equation}\label{Gibbs1}
TdS_D=dE_D+p_DdV.
\end{equation}
In this equation $S_D$ is the entropy of dark energy candidate,
$V_D=\frac{4}{3}\pi \tilde{r_A}$ and $E_D=\rho_DV$ are the volume
of the flat FRW universe and the energy of dark energy,
respectively. The temperature of fluids confined by the apparent
horizon is equal to the Cai-Kim temperature
\cite{CaiKim,sheyw2,Cai2}
\begin{equation}\label{temperature00}
T=\frac{1}{2\pi\tilde{r}_A}.
\end{equation}
Bearing Eq.~(\ref{ah}) in mind, this equation leads to
\begin{equation}\label{temperature0}
T=\frac{H}{2\pi},
\end{equation}
for a flat background ($k=0$). Since
\begin{equation}\label{dED}
dE_D=\rho_DdV+Vd\rho_D,
\end{equation}
and
\begin{equation}\label{dv_D}
dV=4\pi(\tilde{r_A})^2d\tilde{r_A}=-4\pi H^{-4}dH,
\end{equation}
Eq.~(\ref{Gibbs1}) becomes
\begin{equation}\label{Gibbs2}
dS_D=\frac{2\pi}{H}\left((\rho_D+p_D)dV+Vd\rho_D\right),
\end{equation}
which yields
\begin{equation}\label{ds0_dH0}
\frac{dS_D^0}{dH_0}=\frac{2
\pi}{H_0^3}(1+K(\rho_D^0)^{1/n})\{-\frac{3}{2}\Omega_D^0+\frac{1}{\frac{1}{\Omega_D^0}+K(\rho_D^0)^{1/n}
}\},
\end{equation}
where we have used Eq.~(\ref{EM2}) to get the above equation and
we have also used the following equation
\begin{equation}\label{rodat2_0}
\frac{d\rho_D^0}{dH_0}=\frac{-3H_0(\rho_D^0+K(\rho_D^0)^{1+\frac{1}{n}})}{-4\pi(1+u^0+K(\rho_D^0)^{\frac{1}{n}})}.
\end{equation}
In order to obtain this equation, use Eqs.~(\ref{friedman_eq})
and~(\ref{EM2}) to get
\begin{equation}\label{rodat2_01}
\dot H_0=-4\pi\rho_D^0(1+u^0+K(\rho_D^0)^{\frac{1}{n}})
\end{equation}
which is Rechaudhury equation and
\begin{equation}\label{rodat2_02}
\dot \rho_D^0=-3H_0\rho_D^0(\rho_D^0+K(\rho_D^0)^{1+\frac{1}{n}}),
\end{equation}
respectively. Now, by inserting these results into the
\begin{equation}\label{rodat_0}
d\rho_D^0=\frac{\dot \rho_D^0 }{\dot H_0}dH_0
\end{equation}
equation one can reach Eq.~(\ref{rodat2_0}). In the above
formulas, the subscript/superscript ($0$) points to a universe in
which dark sectors do not interact with each other. Loosely
speaking, Eq.~(\ref{ds0_dH0}) gives us a relation for the entropy
changes of dark energy candidate, satisfying the Polytropic
equation of state~(\ref{EOS}), in the absence of an interaction
with dark matter candidate.
\subsection{Thermodynamic description of interacting Polytropic gas}
For a universe in which dark sectors interact with each other, the
energy-momentum conservation law leads to
\begin{equation}\label{EM100}
\dot{\rho}_m+3H\rho_m=Q,
\end{equation}
and
\begin{equation}\label{EM200}
\dot{\rho}_D+3H(\rho_D+p_D)=-Q,
\end{equation}
where $Q$ denotes the mutual interaction between the dark sectors of
cosmos. Therefore, $Q>0$ and $Q<0$ mean dark energy is decaying into
dark matter and vice versa, respectively. In this manner, using
Eqs.~(\ref{EM200}) and~(\ref{friedman_eq}) to get
\begin{equation}\label{rodat2}
d\rho_D=\frac{-Q-3H(\rho_D+K(\rho_D)^{1+\frac{1}{n}})}{-4\pi(1+u+K\rho_D^{\frac{1}{n}})}dH.
\end{equation}
Now, following approach which leads to Eq.~(\ref{ds0_dH0}), to
reach
\begin{equation}\label{ds_dH3}
\frac{dS_D}{dH}=\frac{2\pi}{H^3}(1+K\rho_D^{\frac{1}{n}})\left[
-\frac{3}{2}\Omega_D +
\frac{1}{\frac{1}{\Omega_D}+K\rho_D^{\frac{1}{n}}}+\frac{Q}
{3H(\frac{1}{\Omega_D}+K\rho_D^{\frac{1}{n}})} \right],
\end{equation}
as a relation for the entropy changes in an interacting universe.
It is apparent that, as an appropriate limit, in the absence of
interaction term ($Q$), this relation converges to the previous
result obtained in Eq.~(\ref{ds0_dH0}). It is also useful to note
here that in such a universe the Hubble parameter is $H$ which
differs from the Hubble parameter of the previous non-interacting
universe ($H_0$). Since an interaction between dark sectors is allowed
from observational point of view which may lead to solve the coincidence
problem \cite{z1,z2,z3,pavonz,z4,z5,z6,z7,z8,wangpavon,z9,co1,z10,co2,co3},
various interactions between the Polytropic model of dark energy and dark
matter has been studied in vast articles \cite{taji,ansari,karami2010,ref59,ahep}.
\subsection{Thermodynamic description of mutual interaction}
In this subsection, we want to find a thermodynamical description
for a mutual interaction between the dark sectors of cosmos. In
fact, a weak interaction may leave fluctuations into the thermal
properties of systems \cite{landau}. Due to this hypothesis, the
entropy of interacting dark energy ($S_D$) can be expressed as the
non-interacting system parameters as
\cite{wangpavon,fl1,fl2,fl3,fl4,fl5,em}:
\begin{equation}\label{SD_total}
S_D=S_D^{0}+S_D^{1}+S_D^{2}
\end{equation}
where $S_D^{0}$ is the entropy of non-interacting dark energy
candidate which obeys Eq.~(\ref{ds0_dH0}). Moreover, $S_D^{1}$ is
evaluated as
\begin{equation}\label{Sd_1}
S_D^{1}=-\frac{1}{2}\ln CT_0^2,
\end{equation}
in which
\begin{equation}\label{T0}
T_0=\frac{H_0}{2\pi}
\end{equation}
and
\begin{equation}\label{capacity1}
C=T_0\frac{\partial S_D^{0} }{\partial T_0}=2\pi
T_0\frac{dS_D^{0}}{dH_0},
\end{equation}
are the temperature and dimensionless heat capacity of
non-interacting dark energy candidate, respectively. $S_D^{2}$
also includes higher order fluctuations
\cite{landau,das,wangpavon,fl1,fl2,fl3,fl4,fl5,em}. After some
algebra we get
\begin{equation}\label{Capacity2}
C=\frac{2\pi}{4\pi^2T_0^2}\left(1+K(\rho_D)^{\frac{1}{n}}\right)\left\lbrace
-\frac{3}{2}\Omega_D^0+\frac{1}{\frac{1}{\Omega_D^0}+K(\rho_D^0)^{\frac{1}{n}}}
\right\rbrace,
\end{equation}
which leads to
\begin{equation}\label{SD_1b}
S_D^{1}=-\frac{1}{2}\ln \left[
\frac{1}{2\pi}\left(1+K(\rho_D^0)^{\frac{1}{n}}\right)
\left\lbrace -\frac{3}{2}\Omega_D^0
+\frac{1}{\frac{1}{\Omega_D^0}+K(\rho_D^0)^{\frac{1}{n} }
}\right\rbrace \right],
\end{equation}
and
\begin{eqnarray}\label{dsd1_dh0}
&&\frac{dS_D^{1}}{dH_0}=\frac{dS_D^{(1)}}{d\rho_D^0}\frac{d\rho_D^0}{dH_0}\\
\nonumber &&=-\frac{1}{2}\left[
\frac{K(\rho_D^0)^{\frac{1}{n}-1}}{n\left(1+K(\rho_D^0)^{\frac{1}{n}}\right)}-\frac{3}{2\rho_c
\left\lbrace -\frac{3}{2}\Omega_D^0
+\frac{\rho_D^0}{\frac{1}{\Omega_D^0}+K(\rho_D^0)^{\frac{1}{n}}
}\right\rbrace}
-\frac{\frac{1}{\Omega_D^0}-\frac{K}{n}(\rho_D^0)^{\frac{1}{n}}}{\left[
\rho_D^{0}\left(1+K(\rho_D^0)^{\frac{1}{n}}\right)^2 \left\lbrace
-\frac{3}{2}\Omega_D^0
+\frac{\rho_D^0}{\frac{1}{\Omega_D^0}+K(\rho_D^0)^{\frac{1}{n} }
}\right\rbrace \right]} \right]\frac{d\rho_D^0}{dH_0}
\end{eqnarray}
where we have used Eqs.~(\ref{capacity1}) and~(\ref{Sd_1}) to
obtain these relations. It is also useful to mention here that
$\frac{d\rho_D^0}{dH_0}$ can be found in Eq.~(\ref{rodat2_0}).
Now, using Eq.~(\ref{SD_total}) to reach
\begin{equation}\label{SD2_2}
\frac{dS_D^{2}}{dH_0}=\frac{dS_D}{dH}\frac{dH}{dH_0}-\frac{dS_D^0}{dH_0}-\frac{dS_D^1}{dH_0},
\end{equation}
in which $\frac{dH_0}{dH}$ can be evaluated as
\begin{equation}\label{dH0_dh}
\frac{dH_0}{dH}=\frac{\dot H_0}{\dot
H}=\frac{\rho_D^0(1+u^0+K(\rho_D^0)^{\frac{1}{n}})}{\rho_D(1+u+K\rho_D^{\frac{1}{n}})}.
\end{equation}
Finally, by inserting Eqs.~(\ref{ds0_dH0}),~(\ref{ds_dH3})
and~(\ref{dsd1_dh0}) into Eq.~(\ref{SD2_2}) and using
Eq.~(\ref{rodat2_0}) together with Eq.~(\ref{dH0_dh}), one can
find a relation for the $S_D^2$ term. Therefore, fluctuation
theorem helps us establish a relation between thermal fluctuations
of dark energy entropy and the mutual interaction between the dark
sides of cosmos. Since the $S_D^2$ term is negligible, one may
neglect from this term and write
\cite{landau,das,wangpavon,fl1,fl2,fl3,fl4,fl5,em}
\begin{equation}\label{SD2_20}
\frac{dS_D}{dH}\approx(\frac{dS_D^0}{dH_0}+\frac{dS_D^1}{dH_0})\frac{dH_0}{dH},
\end{equation}
in which $\frac{dH_0}{dH}$ follows Eq.~(\ref{dH0_dh}). Now, by
inserting Eqs.~(\ref{ds0_dH0}),~(\ref{ds_dH3})
and~(\ref{dsd1_dh0}) into this equation one may find a relation
for $Q$ up to the first order of fluctuations. Roughly speaking,
this relation helps us get a thermodynamic interpretation for the
mutual interaction between the dark sectors of cosmos, up to the
first order of fluctuations.
\section{Thermodynamic description of mutual interaction between viscous Polytropic gas and dark matter}
In this section we generalize our debates, expressed in the
previous section, to a viscous Polytropic gas which satisfies the
\begin{equation}\label{EOS_viscous}
p_D=K\rho_D^{1+\frac{1}{n}}-3\xi H,
\end{equation}
equation of state, where $\xi$ is viscous coefficient \cite{ahep}. Bearing this equation as well as
the Friedman equation~(\ref{friedman_eq}) in mind, since for a
non-interacting viscous Polytropic gas Eqs.~(\ref{EM1})
and~(\ref{EM2}) are valid, by following the approaches yield
Eqs.~(\ref{ds0_dH0}) and~(\ref{rodat2_0}) one gets
\begin{equation}\label{ds0_viscose}
dS_D^0=\frac{2\pi}{H_0^3}\left\lbrace
\frac{1+K(\rho_D^0)^{\frac{1}{n}}-\frac{-3\xi
H_0}{\rho_D^0}}{(\Omega_D^0)^{-1}+K(\rho_D^0)^{\frac{1}{n}}-\frac{3\xi
H_0}{\rho_D^0}}-\frac{3}{2}\Omega_D^0\left((1+K(\rho_D^0)^{\frac{1}{n}})+\frac{3\xi
H_0}{\rho_D^0} \right) \right\rbrace dH_0,
\end{equation}
and
\begin{equation}\label{dro_0_viscose}
\frac{d\rho_D^0}{dH_0}=\frac{-3H_0\left\lbrace
1+K(\rho_D^0)^{\frac{1}{n}}-\frac{3\xi
H_0}{\rho_D^0}\right\rbrace}{-4\pi \left\lbrace
\frac{1}{\Omega_D^0}+K(\rho_D^0)^{\frac{1}{n}}-\frac{3\xi
H_0}{\rho_D^0} \right\rbrace},
\end{equation}
respectively. It is obvious that these equations cover previous
results (Eqs.~(\ref{ds0_dH0}) and~(\ref{rodat2_0})), in the
appropriate limit $\xi\rightarrow0$.
\subsection*{Thermodynamic description of interacting viscous Polytropic gas}
When there is an interaction, we have \cite{ahep}
\begin{equation}\label{EM1000}
\dot{\rho}_m+3H\rho_m=Q,
\end{equation}
and
\begin{equation}\label{EM2000}
\dot{\rho}_D+3H(\rho_D+p_D)=-Q,
\end{equation}
where $Q$ again denotes the mutual interaction between the dark
sectors of cosmos. Now, by using Eqs.~(\ref{friedman_eq})
and~(\ref{EOS_viscous}) we get
\begin{equation}\label{dotH_viscose}
\dot H=-4 \pi \rho_D \left(1+u+K\rho_D^{\frac{1}{n}}-\frac{3\xi
H}{\rho_D} \right),
\end{equation}
and
\begin{equation}\label{dotrovisint}
\dot \rho_D=-3H\rho_D\left\lbrace
1+K(\rho_D)^{\frac{1}{n}}-\frac{3\xi H}{\rho_D}\right\rbrace-Q,
\end{equation}
respectively. By combining these equations with each other, one
gets
\begin{equation}\label{dro_dhviscoseint}
\frac{d\rho_D}{dH}=\frac{3H\rho_D\left\lbrace
1+K(\rho_D)^{\frac{1}{n}}-\frac{3\xi
H}{\rho_D}\right\rbrace+Q}{4\pi \rho_D\left\lbrace
\frac{1}{\Omega_D}+K(\rho_D)^{\frac{1}{n}}-\frac{3\xi H}{\rho_D}.
\right\rbrace}
\end{equation}
Bearing the approach which leads to Eq.~(\ref{ds0_dH0}) in mind,
by some calculations we get
\begin{eqnarray}\label{ds_dh_final}
\frac{dS_D}{dH}&=& \frac{{2\pi}}{{{H^3}}}\left[ {\frac{{1 +
K{{({\rho _D})}^{\frac{1}{n}}} - \frac{{3\xi H}}{{{\rho
_D}}}}}{{{{({\Omega _D})}^{ - 1}} + K{{({\rho _D})}^{\frac{1}{n}}}
- \frac{{3\xi H}}{{{\rho _D}}}}}} \right. - \frac{3}{2}{\Omega
_D}\left( {1 + K{{({\rho _D})}^{\frac{1}{n}}} - \frac{{3\xi
H}}{{{\rho _D}}}} \right)\\ \nonumber &&\left.{ +
\frac{Q}{{3H{\rho _D}\left( {1 + u + K{{({\rho
_D})}^{\frac{1}{n}}} - \frac{{3\xi H}}{{{\rho _D}}}}
\right)}(1+K(\rho_D)^{\frac{1}{n}})}} \right],
\end{eqnarray}
as a relation for the entropy changes of an interacting viscous
Polytropic dark energy model confined by the apparent horizon. It
is obvious that in the $Q\rightarrow 0$ limit
Eq.~(\ref{ds0_viscose}) is covered, whiles Eq.~(\ref{ds0_dH0}) is
govern by taking the $Q\rightarrow 0$ and $\xi\rightarrow0$
limits simultaneously. It is also useful to mention that the
result of non-viscous interacting case~(\ref{ds_dH3}) is
obtainable by imposing the $\xi\rightarrow0$ limit to this
equation. Using Eqs.~(\ref{capacity1}) and~(\ref{Sd_1}) to get
\begin{equation}\label{capacity_2}
C=\frac{1}{{2\pi T_0^2}}\left\{ {\frac{{1 + K{{(\rho
_D^0)}^{\frac{1}{n}}} - \frac{{3\xi {H_0}}}{{\rho
_D^0}}}}{{{{(\Omega _D^0)}^{ - 1}} + K{{(\rho
_D^0)}^{\frac{1}{n}}} - \frac{{3\xi {H_0}}}{{\rho _D^0}}}} -
\frac{3}{2}\Omega _D^0\left( {1 + K{{(\rho _D^0)}^{\frac{1}{n}}} -
\frac{{3\xi {H_0}}}{{\rho _D^0}}} \right)} \right\},
\end{equation}
and
\begin{equation}\label{SD_1viscos}
S_D^1=-\frac{1}{2}\ln\left\{ {\frac{1}{{2\pi }}\left( {\frac{{1 +
K{{(\rho _D^0)}^{\frac{1}{n}}} - \frac{{3\xi {H_0}}}{{\rho
_D^0}}}}{{{{(\Omega _D^0)}^{ - 1}} + K{{(\rho
_D^0)}^{\frac{1}{n}}} - \frac{{3\xi {H_0}}}{{\rho _D^0}}}} -
\frac{3}{2}\Omega _D^0\left( {1 + K{{(\rho _D^0)}^{\frac{1}{n}}} -
\frac{{3\xi {H_0}}}{{\rho _D^0}}} \right)} \right)} \right\},
\end{equation}
for the dimensionless heat capacity and the first order term of
thermal fluctuations of dark energy candidate, respectively. The
latter leads to
\begin{eqnarray}\label{ds1_dho_viscose} \frac{dS_D^1}{dH_0}&=&
\frac{{d\rho _D^0}}{{d{H_0}}}\times \frac{-1}{2} \times \\
\nonumber &&\left\{\frac{{\left( {\frac{K}{n}{{(\rho
_D^0)}^{\frac{1}{n} - 1}} + \frac{{3\xi {H_0}}}{{{{(\rho
_D^0)}^2}}}} \right)\left( {\frac{1}{{\Omega _D^0}} + K{{(\rho
_D^0)}^{\frac{1}{n}}} - \frac{{3\xi {H_0}}}{{\rho _D^0}}}
\right)+\left( { \frac{{{\rho _c}}}{{{{(\rho _D^0)}^2}}} -
\frac{K}{n}{{(\rho _D^0)}^{\frac{1}{n} - 1}} - \frac{{3\xi
{H_0}}}{{{{(\rho _D^0)}^2}}}} \right)\left( {1 + K{{(\rho
_D^0)}^{\frac{1}{n}}} - \frac{{3\xi {H_0}}}{{\rho _D^0}}}
\right)}}{{\frac{{1 + K{{(\rho _D^0)}^{\frac{1}{n}}} - \frac{{3\xi
{H_0}}}{{\rho _D^0}}}}{{\frac{1}{{\Omega _D^0}} + K{{(\rho
_D^0)}^{\frac{1}{n}}} - \frac{{3\xi {H_0}}}{{\rho _D^0}}}} -
\frac{3}{2}\Omega _D^0\left( {1 + K{{(\rho _D^0)}^{\frac{1}{n}}} -
\frac{{3\xi {H_0}}}{{\rho _D^0}}} \right)}} \right.\\ \nonumber &&
\left. - \frac{3}{2}\frac{{\frac{1}{{{\rho _0}}}\left( {1 +
K{{(\rho _D^0)}^{\frac{1}{n}}} - \frac{{3\xi {H_0}}}{{\rho _D^0}}}
\right) + \Omega _D^0\left( {\frac{K}{n}{{(\rho
_D^0)}^{\frac{1}{n} - 1}} + \frac{{3\xi {H_0}}}{{{{(\rho
_D^0)}^2}}}} \right)}}{{\frac{{1 + K{{(\rho _D^0)}^{\frac{1}{n}}}
- \frac{{3\xi {H_0}}}{{\rho _D^0}}}}{{\frac{1}{{\Omega _D^0}} +
K{{(\rho _D^0)}^{\frac{1}{n}}} - \frac{{3\xi {H_0}}}{{\rho
_D^0}}}} - \frac{3}{2}\Omega _D^0\left( {1 + K{{(\rho
_D^0)}^{\frac{1}{n}}} - \frac{{3\xi {H_0}}}{{\rho _D^0}}}
\right)}}\right\},
\end{eqnarray}
where $\frac{d\rho_D^0}{dH_0}$ can be found in
Eq.~(\ref{dro_0_viscose}). Bearing the thermal fluctuations
theory~(\ref{SD_total}) in mind, by using
Eqs.~(\ref{ds1_dho_viscose}),~(\ref{ds_dh_final}) together
with~(\ref{ds0_viscose}) we get \cite{emint}
\begin{equation}\label{SD2_eq}
\frac{dS_D^2}{dH_0}=\frac{dS_D}{dH_0}\frac{dH_0}{dH}-\frac{dS_D}{dH_0}-\frac{dS_D^1}{dH_0},
\end{equation}
where
\begin{equation}\label{dH0_dh_final}
\frac{dH_0}{dH}=\frac{\dot{H}_0}{\dot{H}}=\frac{\rho_D^0}{\rho_D}\frac{(1+u^0+K(\rho_D^0)^{\frac{1}{n}})-\frac{3\xi
H_0}{\rho_D^0}}{(1+u+K\rho_D^{\frac{1}{n}})-\frac{3\xi
H}{\rho_D}},
\end{equation}
for the higher order terms of thermal fluctuations of the dark
energy candidate entropy and an expression for the relation between
the Hubble parameters in the interacting and non-interacting
universes, respectively. Moreover, if one only considers the $S_D^0$
and $S_D^1$ terms she can find a relation for the mutual interaction
between dark sectors up to the first order of thermal fluctuations,
by inserting
Eqs.~(\ref{ds1_dho_viscose}),~(\ref{ds_dh_final}),~(\ref{ds0_viscose})
and~(\ref{dH0_dh_final}) into relation
\cite{fl1,fl2,fl3,fl4,fl5,emint}
\begin{eqnarray}\label{SD2_eq0}
\frac{dS}{dH}\approx(\frac{dS_D}{dH_0}+\frac{dS_D^1}{dH_0})\frac{dH_0}{dH}.
\end{eqnarray}
In the above equations, the subscript and superscript ($0$) point
to the non-interacting viscous case. Therefore, by relating the
mutual interaction between dark sectors of cosmos to the thermal
fluctuation theory, we could find a relation for the mutual
interaction up to the first order of fluctuations together with a
relation for the higher order of fluctuations ($S_D^2$).
\subsection*{Coincidence problem and mutual interaction}
The most general form of the mutual interaction between the dark
energy candidate, either satisfying the Polytropic equation
state~(\ref{EOS}) or its viscous form~(\ref{EOS_viscous}), and the
pressureless dark matter can be written as
\cite{taji,ansari,karami2010,ref59,ahep}
\begin{eqnarray}\label{int}
Q=3b_1H\rho_D+3b_2H\rho_m,
\end{eqnarray}
where $b_1$ and $b_2$ are coupling constants of interaction. In
order to solve the coincidence problem, interaction term should
satisfy the $Q>0$ condition \cite{fl5,emint} which leads to
\begin{eqnarray}\label{inttt}
\frac{b_1}{b_2}>-\frac{1-\Omega_D}{\Omega_D}=-u,
\end{eqnarray}
where we have used the $\Omega_i$ and $u$ definitions together with
Eq.~(\ref{eq_u}) to obtain this equation. It is worth to mention
here that since in the current era of universe expansion the
density of dark energy is approximately $3$ times larger than that
of dark matter \cite{roos}, $u\approx\frac{1}{3}$ which means that
the $b_1>-\frac{b_2}{3}$ condition should be met by the
interaction coupling constants.
\section{Dark energy candidate may modify the Horizon entropy}
Here, we are going to study the thermodynamics of apparent horizon
in a universe in which the dark energy candidate obeys either the
Polytropic equation of state~(\ref{EOS}) or its viscous
form~(\ref{EOS_viscous}). In order to achieve this goal, we try to
find an expression for the horizon entropy by bearing in mind this
fact that since the nature of DE candidate may be not similar to
other parts of cosmos, it may affect the horizon entropy
\cite{cana,cana1,em,mmg,md}. We also use the Cai-Kim approach in
which the volume changes of universe in the infinitesimal time
$dt$ can be neglected ($dV\approx0$) \cite{CaiKim}. Finally, we
point to the second law of thermodynamics and thus a required
condition for availability of the second law of thermodynamics
\cite{haw}.

\subsection{Non-interacting and non-viscous Polytropic model}\label{entropymodify1}

Since we want to calculate the entropy of horizon, we take into
account the various parts of cosmos constitutes. Therefore,
consider a FRW universe with Friedman equation
\begin{eqnarray}\label{Friedmann1}
\frac{1}{\tilde{r}_A^2}= \frac{8\pi}{3}(\rho+\rho_d),
\end{eqnarray}
in which we have used Eq.~(\ref{ah}), and
\begin{eqnarray}\label{oden}
\rho=\rho_{cdm}+\rho_{wdm}+...
\end{eqnarray}
includes the energy density of everything filling the universe,
such as the cold and worm dark matters and etc.., except the dark
energy candidate. In the non-interacting universe the
energy-momentum conservation law implies
\begin{eqnarray}\label{ocon1}
\dot{\rho}+3H(\rho+p)=0,
\end{eqnarray}
and
\begin{eqnarray}\label{dcon1}
\dot{\rho}_{D}+3H(\rho_{D}+p_{D})=0,
\end{eqnarray}
where $p$ and $p_D$ denote the corresponding pressure. The
projection of total energy-momentum tensor ($T_{\mu \nu}$) on the
normal direction of the two-dimensional sphere with radii $\zeta$
and the energy flux crossing this sphere read as \cite{Cai2}
\begin{eqnarray}\label{esv}
\psi_a = T^b_a\partial_b \zeta + W\partial_a \zeta,
\end{eqnarray}
and
\begin{eqnarray}\label{uf4}
\delta Q^m=-AH\zeta(\frac{\rho+p}{2})dt+Aa(\frac{\rho+p}{2})dr,
\end{eqnarray}
respectively. In this equation, $\delta Q^m$ denotes the energy
amount crossing two-dimensional hypersurface with radii $\zeta$
during the universe expansion. Since $\zeta=ar$, the latter can be
rewritten as
\begin{eqnarray}\label{ufl1}
\delta
Q^m=-\frac{3V(\rho+p)H}{2}dt+\frac{A(\rho+p)}{2}(d\zeta-\zeta H
dt).
\end{eqnarray}
In the Cai-Kim approach to get the horizon entropy \cite{CaiKim},
the temperature of fluids confined by the apparent horizon reads
Eq.~(\ref{temperature00}) and the $d\zeta \approx 0$ approximation
is used in the infinitesimal time $dt$. Therefore,
Eq.~(\ref{ufl1}) takes the
\begin{eqnarray}\label{ufl2}
\delta Q^m=-(\frac{3V}{2}+\frac{A\zeta}{2})(H(\rho+p)dt),
\end{eqnarray}
form. Moreover, since $A\zeta=3V$, we obtain
\begin{eqnarray}\label{ufl3}
\delta Q^m=-3VH(\rho+p)dt=Vd\rho,
\end{eqnarray}
where we have used Eq.~(\ref{ocon1}) to get the last equality.
Bearing the Clausius relation in mind \cite{CaiKim}, we get
\begin{eqnarray}\label{claus1}
dS_A\equiv -\frac{\delta Q^m}{T}=-\frac{V}{T}d\rho.
\end{eqnarray}
Now, taking differentiation from Eq.~(\ref{Friedmann1}), and
inserting the results into this equation to reach
\begin{eqnarray}\label{general}
dS_A=(2\pi \tilde{r}_A+\frac{8\pi^2}{3}\tilde{r}_A^4\rho_D^{\prime})d\tilde{r}_A,
\end{eqnarray}
where $\rho_D^{\prime}=\frac{d\rho_D}{d\tilde{r}_A}$. After integration we
get
\begin{eqnarray}\label{ent1}
S_A=\frac{A}{4}-8\pi^2 \int
\frac{H\rho_D(1+K\rho_D^{\frac1n})}{(H^2+\frac{k}{a^2})^2}dt,
\end{eqnarray}
where we have used Eq.~(\ref{ah}) and $A=4\pi\tilde{r}_A^2$
together with Eqs.~(\ref{EOS}) and~(\ref{dcon1}) to obtain this
equation. We have also set the integration constant ($S_0$) to
zero. This equation is in full agreement with previous study about
the effect of a dark energy candidate with varying density profile
on the horizon entropy in the FLRW universe with curvature $k$
\cite{md}. For a flat background ($k=0$), this equation is
moderated to
\begin{eqnarray}\label{ent12}
S_A=\frac{A}{4}-8\pi^2 \int
\frac{\rho_D(1+K\rho_D^{\frac1n})}{H^3}dt,
\end{eqnarray}
which is in full agreement with previous studies about the effects
of dark energy candidate with varying energy density on the
horizon entropy in a flat background \cite{em}.
\subsection*{Second law of thermodynamics}
The second law of thermodynamics states that the horizon entropy
as the total entropy of a gravitational system should obey the
$\frac{dS_A}{dt}\geq0$ condition \cite{haw}. By combining
Eqs.~(\ref{claus1}) and~(\ref{ufl3}), we reach
\begin{eqnarray}\label{claus1}
\frac{dS_A}{dt}=\frac{3VH(\rho+p)}{T}.
\end{eqnarray}
Therefore the second law of thermodynamics is satisfied whenever,
the $\rho+p\geq0$ condition is met. The same result is previously
reported for some other dark energy candidates \cite{em,md}.
\subsection{Interacting non-viscous Polytropic model}\label{interacting}
Bearing Eqs.~(\ref{Friedmann1}) and~(\ref{oden}) in mind, whiles
the cosmos sectors interact with each other, the energy-momentum
conservation law is written as
\begin{eqnarray}\label{ocon11}
\dot{\rho}+3H(\rho+p)=Q,
\end{eqnarray}
and
\begin{eqnarray}\label{dcon11}
\dot{\rho}_{D}+3H(\rho_{D}+p_{D})=-Q.
\end{eqnarray}
Now, taking into account the approach of
section~(\ref{entropymodify1}) and using~(\ref{ocon11}) to get
\begin{eqnarray}\label{ufl33}
\delta Q^m=V(d\rho-Qdt),
\end{eqnarray}
where $\delta Q^m$ is again the energy amount crossing the boundary.
Finally, using Eq.~(\ref{Friedmann1}) together with Clausius
equation ($TdS_A=-\delta Q^m$) to get
\begin{eqnarray}\label{claus11}
dS_A=2\pi \tilde{r}_A d\tilde{r}_A
+\frac{8\pi^2}{3}\tilde{r}_A^4(d\rho_D+Qdt),
\end{eqnarray}
which leads to
\begin{eqnarray}\label{claus111}
S_A=\frac{A}{4} +\frac{8\pi^2}{3}\int \tilde{r}_A^4(d\rho_D+Qdt).
\end{eqnarray}
This equation shows that how the mutual interaction between the
Polytropic gas ( as the dark energy candidate) and other parts of
cosmos affects the horizon entropy, which is fully consistent with
previous studies \cite{em,md}. By using Eq.~(\ref{dcon11}), this
equation can be rewritten as
\begin{eqnarray}
S_A=\frac{A}{4}-8\pi^2 \int
\frac{H\rho_D(1+K\rho_D^{\frac1n})}{(H^2+\frac{k}{a^2})^2}dt,
\end{eqnarray}
which is similar to the previous result~(\ref{ent1}). Indeed, we
should note here that although this equation is similar to
Eq.~(\ref{ent1}) but due to the interaction term ($Q$) they differ
from each other. The identical result is obtained for other models of dark energy \cite{em,md}.
\subsection*{Second law of thermodynamics}
In order to investigate the validity of second law of
thermodynamics, combine Eqs.~(\ref{ocon11}),~(\ref{ufl33}) and
Clausius relation to obtain
\begin{eqnarray}
\frac{dS_A}{dt}=\frac{3HV}{T}(\rho+p),
\end{eqnarray}
which claims that, just the same as the non-interacting case, the
second law of thermodynamics is met if the $\rho+p\geq0$ condition
is satisfied. This result is again in accordance with previous studies about other models of dark energy \cite{em,md}.
\subsection{Non-interacting viscous Polytropic model}
In order to get an expression for the horizon entropy, since dark
energy candidate does not interact with other parts of cosmos,
following the approach of Sec.~(\ref{entropymodify1}) to reach
\begin{eqnarray}\label{claus2}
dS_A=2\pi \tilde{r}_A d\tilde{r}_A
+\frac{8\pi^2}{3}\tilde{r}_A^4d\rho_D.
\end{eqnarray}
It is useful to note here that although, since there is no
interaction between the cosmos sectors, this result is similar to
Eq.~(\ref{claus11}), but due to the viscosity, the result of this
equation differs from that of~(\ref{claus11}). In order to make the
difference between this equation and~(\ref{claus11}) clear and find
the viscosity effect on the horizon entropy, bearing
Eq.~(\ref{EOS_viscous}) in mind, and using Eq.~(\ref{dcon1}) to
get
\begin{eqnarray}\label{claus3}
S_A=\frac{A}{4}-8\pi^2 \int
\frac{H\rho_D(1+K\rho_D^{\frac1n})}{(H^2+\frac{k}{a^2})^2}dt-24\pi^2\int
\frac{\xi H^2}{(H^2+\frac{k}{a^2})^2}dt.
\end{eqnarray}
Therefore, the third term at the RHS of this equation is a
criterion for the effect of the viscosity of dark energy candidate
on the horizon entropy. Since the third term at the RHS of this equation depends only on the viscosity coefficient $\xi$ and the spacetime parameter, including the Hubble parameter, scale factor and the curvature constant $k$, this term is identical for all dark energy candidates with the viscosity modification $-3\xi H$ to the pressure of dark energy candidate. It is apparent that, in the $\xi\rightarrow 0$ limit, the result of Sec.~(\ref{entropymodify1}), as a desired result, is obtainable. It is also easy to check that the second law of thermodynamics is satisfied if the $\rho+p\geq0$ condition is met.
\subsection{Interacting viscous Polytropic model}
In the interacting cosmos case, by following the approach used in
Sec.~(\ref{interacting}) we get
\begin{eqnarray}
dS_A=2\pi \tilde{r}_A d\tilde{r}_A
+\frac{8\pi^2}{3}\tilde{r}_A^4(d\rho_D+Qdt),
\end{eqnarray}
which leads to
\begin{eqnarray}\label{vis31}
S_A=\frac{A}{4}-8\pi^2 \int
\frac{H\rho_D(1+K\rho_D^{\frac1n})}{(H^2+\frac{k}{a^2})^2}dt-24\pi^2\int
\frac{\xi H^2}{(H^2+\frac{k}{a^2})^2}dt,
\end{eqnarray}
where we have used Eqs.~(\ref{EOS_viscous}) and~(\ref{EM2000}) to
obtain this equation. It is useful to note here that although this
equation is similar to Eq.~(\ref{claus3}), but they are completely
differ from each other. It is due to this fact that due to the
interaction $Q$, the profile density ($\rho_D$) and pressure ($p_D$)
of dark energy candidate differ from their values in a
non-interacting universe ($Q=0$). It is also easy to check that the
second law of thermodynamics is met whenever the $\rho+p\geq 0$
condition is satisfied. Finally, we should mention that the results
of non-viscous and non-interacting cosmoses are recovered by
inserting the $\xi=0$ and $Q=0$ conditions, respectively.

\section{Summary and concluding remarks}
Throughout this paper, we tried to investigate the various
thermodynamical aspects of a dark energy candidate which either
satisfies the Polytropic equation of state~(\ref{EOS}) or its
viscous type~(\ref{EOS_viscous}). We also adopted the Cai-Kim
temperature~(\ref{temperature0}) to the horizon and fields which
fill the universe \cite{CaiKim}. Moreover, since the major part of
cosmos is dark and the current dynamics of universe is completely
compatible with its dark sectors contents, we omit the
contribution of non-dark parts to the Friedmann equation in the
second and third chapters. In section ($\textmd{II}$), we
considered a situation in which dark energy candidate, enclosed by
apparent horizon, satisfies the Polytropic equation of
state~(\ref{EOS}) as well as the first law of thermodynamics. In
continue, we could get a relation for the entropy changes of the
dark energy candidate~(\ref{ds0_dH0}), whiles dark sectors do not
interact with each other. In addition, we generalized our
calculations to a case in which there is a mutual interaction
between dark sectors, and found a relation for the entropy
changes~(\ref{ds_dH3}). Thereinafter, by using the thermal
fluctuation theorem~(\ref{SD_total}), we could get a relation for
higher order thermal fluctuations of the entropy of dark energy
candidate~(\ref{SD2_2}). Finally, we showed that if one neglects
the higher order terms, she can get a relation for the mutual
interaction between the dark sectors of cosmos based on the
thermal fluctuations of dark energy candidate~(\ref{SD2_20}).
Indeed, we established a relation between the mutual interaction
between the dark sectors of cosmos and thermal fluctuations of
dark energy, which leads to a thermodynamical interpretation for
the mutual interaction between the dark sectors of cosmos, meaning
that one may find a relation for such interaction by following the
thermal history of dark energy candidate. In section
($\textmd{III}$), we have generalized our debates to a universe in
which dark energy candidate satisfies the viscous Polytropic
equation of state~(\ref{EOS_viscous}), and followed the recipe
used in the second section to get the thermodynamical description
for the non-interacting viscous Polytropic dark
energy~(\ref{ds0_viscose}), interacting viscous Polytropic dark
energy~(\ref{ds_dh_final}) as well as the higher order thermal
fluctuations~(\ref{SD2_eq}) and mutual interaction between the
dark sectors~(\ref{SD2_eq0}). We have also found a relation
between the Hubble parameters in interacting and non-interacting
universes for both of the viscous~(\ref{dH0_dh_final}) and
non-viscous Polytropic~(\ref{dH0_dh}) models of dark energy.
Finally, we have considered the general form of mutual interaction
between dark sectors and got a condition which should be satisfied
by the coupling constant of interaction term in order to alleviate
the coincidence problem~(\ref{inttt}). In the forth chapter, we
have taken into account the view in which the dark energy
candidate may modify the horizon entropy
\cite{cana,cana1,em,mmg,md} as well as the Cai-Kim approaches
\cite{CaiKim} to get the Friedmann equations by applying the
thermodynamics laws on the apparent horizon. Since all of the
baryonic contents of cosmos may have contribution to the total
entropy and thus the horizon entropy, unlike the second and third
sections, we have tired to consider all of the possible sources,
which fill the cosmos, in the Friedmann equation in our study in
section four. Our study shows that horizon entropy may be modified
by dark energy candidate and its interaction with other parts of
cosmos. As it is clear from our calculations, the contribution of
dark energy candidate to the horizon entropy, irrespective of the
existence or absence of mutual interaction between the dark
sectors, can be written in the $-8\pi^2\int
\frac{H(\rho_D+p_D)}{(H^2+\frac{k}{a^2})^2}dt$ form where the dark
energy candidate either satisfies the Polytropic equation of
state~(\ref{EOS}) or its viscous type~(\ref{EOS_viscous}). Loosely
speaking, it is a term besides the Bekenstein relation for the
horizon entropy, and is in full agreement with the previous
studies including the effects of various dark energy models on the
horizon entropy in vast theories of gravity
\cite{cana,cana1,em,mmg,md}. It is also useful to note here that
the third term in the RHS of Eqs.~(\ref{claus3}) and~(\ref{vis31})
point to the probable effects of viscosity of dark energy
candidate on the horizon entropy in both interacting and
non-interacting cosmoses. Finally, we investigated the validity of
the second law of thermodynamics for the corrected entropy and
find that whenever the $\rho+p\geq0$ condition is met, the second
law of thermodynamics is also obeyed. The latter remark is
independent of the existence or absence of mutual interaction
between the dark sectors and the equation of state satisfied by
the dark energy candidate.

\acknowledgments{The work of H. Moradpour has been supported
financially by Research Institute for Astronomy \& Astrophysics of
Maragha (RIAAM).}

\end{document}